# Confined ionic liquids films under shear: The importance of the chemical nature of the solid surface

Kalil Bernardino,[1*] Mauro C. C. Ribeiro[2]

[1] *Laboratório de Química Teórica*, *Departamento de Química*, *Universidade Federal de São Carlos, Rod. Washington Luiz s/n, 13565-905, São Carlos, Brazil*

[2] *Laboratório de Espectroscopia Molecular*, *Departamento de Química Fundamental*, *Instituto de Química, Universidade de São Paulo, Av. Prof. Lineu Prestes 748, 05508-000, São Paulo, Brazil*

*email: kalilb@ufscar.br

# Confined ionic liquids films under shear: The importance of the chemical nature of the solid surface

Kalil Bernardino,[1*] Mauro C. C. Ribeiro[2]

[1] *Laboratório de Química Teórica*, *Departamento de Química*, *Universidade Federal de São Carlos, Rod. Washington Luiz s/n, 13565-905, São Carlos, Brazil*

[2] *Laboratório de Espectroscopia Molecular*, *Departamento de Química Fundamental*, *Instituto de Química, Universidade de São Paulo, Av. Prof. Lineu Prestes 748, 05508-000, São Paulo, Brazil*

*email: kalilb@ufscar.br

Ionic liquids have generated interest in applications as lubricants and as additives to conventional lubricants due to their unique physical properties. In these applications, the liquid thin film can be subjected simultaneously to extremely high shear and loads in addition to nanoconfinement effects. Here, we use molecular dynamics simulations with a coarse grained model to study a nanometric film of an ionic liquid confined between two planar solid surfaces both at equilibrium and at several shear rates. The strength of the interaction between the solid surface and the ions was changed by simulating three different surfaces with enhanced interactions with different ions. The increase of the interaction with either the cation or the anion leads to the formation of a solid-like layer that moves alongside the substrates, but this layer can exhibit different structures and stability. An increase in interaction with the high symmetry anion produces a more regular structure that is more resistant to the effects of shear and viscous heating. Two definitions were proposed and used for the calculation of the viscosity: a local definition based on microscopic characteristics of the liquid and an engineering definition based on the forces measured at the solid surfaces, with the former displaying a correlation with the layered structure induced by the surfaces. Because of the shear thinning behavior of the ionic liquids as well as the temperature rise brought on by viscous heating, both the engineering and the local viscosities decrease as the shear rate increases.

## 1. INTRODUCTION

Ionic liquids (ILs) are salts with relatively low melting points because of the presence of bulk, flexible and low symmetry ions.[1,2,3] These substances exhibit a number of intriguing physical characteristics, including essentially zero vapor pressure, good surface adhesion, good thermal and chemical stability, and relatively high viscosities when compared to common molecular solvents. They also remain in the liquid phase over a wide range of pressure and temperature. Those properties are desirable for applications as lubricants or as additives to conventional surfactants, where the liquid is simultaneously subject to high temperatures, high loads, and high shear, requiring a characterization of how intermolecular structure and the physical properties, especially the viscosity, change under those conditions.[4,5,6,7,8]

The lubricant coating of an ionic liquid over a solid surface can be as thin as 10 nm.[8,9,10] If a Coeutte flow is produced in the film confined between movable surfaces, even relatively low velocities applied on the covered surface can induce extraordinarily high shear rates. For example, a velocity of only 10 m/s generates a shear rate of $10^9$ $s^{-1}$ = 1 GHz in a 10 nm liquid film, which is orders of magnitude larger than the shear rates of 0.1-1000 $s^{-1}$ that are typically measured in rheology experiments using macroscopic volumes of liquid. Transport properties in confined and interfacial systems can be studied using microrheology methods, which enable the reduction of the liquid volume, but are still limited to dimensions of the order of micrometers instead of nanometers.[11,12] Although challenging for the experiments, this range of thickness and shear rates is appropriate for computer simulation. Two main approaches are used to study the viscosity dependence of bulk liquids using computer simulations: non-equilibrium molecular dynamics (NEMD), where the shear rate is imposed by SLLOD equations of motion and the resulting shear stress is measured,[13] and reverse non-equilibrium molecular dynamics (RNEMD), where the shear stress is controlled by artificial momentum exchanging between particles and the resulting shear rate is measured.[14] The viscosity is obtained in both methods dividing the shear stress to the shear rate. Both NEMD[15,16,17,18] and RNEMD[19,20,21,22] simulations of ionic liquids have demonstrated a shear thinning behavior, *i.e.* decrease of viscosity with increasing shear rate, at rates of GHz that can be achieved in a lubricant film even for liquids that behave like newtonian fluids at lower shear rates.[23]

In addition to the high shear rate, confinement effects in a lubricant film cause the liquid to exhibit radically different intermolecular structure and physical properties from the bulk liquid.[10,24] For instance, computer simulations have demonstrated that water confined in a thin film of 1 nm has a viscosity that is roughly 10 times greater than that of bulk water.[25] In the case of ionic liquids, the surface can promote the preferred adsorption of either the cation or the anion.[26,27] In addition, an ionic liquid near solid surfaces frequently exhibits a layered structure that can reach several nanometers in length.[23,28,29,30] A dramatic impact on the structure and interactions with the confining solid surfaces relies in the fact that the melting point of ionic liquids can both increase[31] or decrease[32] when confined, depending on the interfacial energy between the solid surface and the crystal and the liquid phase.[24]

In this work, we will study simultaneously the effects of the shear rate and nanoconfinement for the ionic liquid 1-butyl-3-methylimidazolium tetrafluoroborate, [BMIM][BF4], by means of molecular dynamics simulations. We are particularly interested in the influence of the intermolecular interactions between the solid surfaces that contain the ionic liquid on its structure and viscosity, both at equilibrium and when subjected to different shear rates. Although molecular dynamics simulations have been used to study ionic liquids thin films interacting with a variety of surfaces, such as rutile,[30] graphene,[33,34] silica,[29,35,36] zeolites[26] and gold,[37,38] only a small number of them have simulated the effects of shear over the confined ionic liquid,[29,34,37] and, to the best of our knowledge, no systematic comparison of the impact of various surfaces over the structure of an ionic liquid under shear has yet been made. This was done here by simulating three different surfaces: the reference surface, which has similar interaction with both ions, a second surface with preferential interaction with the anions, and the third surface, which has preferential interaction with the polar portion of the cation. Additionally, two definitions of viscosity for nanoconfined systems were proposed and computed: A local viscosity that is calculated based on the local properties of the liquid and related to its own dynamics, and an engineering viscosity that is related to the friction force that opposes the movement of the solid surfaces. Details regarding the construction of the model systems and simulation conditions are given in section 2. A detailed discussion of the two viscosity definitions and considerations about viscous heating are given in section 3.1. Sections 3.2 and 3.3 address the effect of the different surfaces on the structure and transport properties of the ionic liquid, respectively.

## 2. METHODOLOGY

In order to assess how the surface's chemical nature and movement speed affect the structure and dynamic properties of an ionic liquid thin film, two model systems were prepared by confining 1050 and 4200 ion pairs between two solid surfaces as shown in Figure 1.a and 1.b, respectively. Due to the large size of the model system and the long timescales involved in the simulations, the coarse-grained force field Martini 3.0[39,40] was used to describe the ionic liquid 1-butyl-3-methylimidazolium tetrafluoroborate, [BMIM][$BF_4$]. Figure 1.c gives the site type attribution, partial charges, and masses. In our previous work,[16] it was demonstrated that this model, which had the relative dielectric constant in Martini force field reduced from 15 to 5, gave a rheological behavior for the bulk [BMIM][BF4] liquid that was similar to the findings from the atomistic and polarizable force field CL&Pol,[41] although with a zero shear viscosity smaller than the experimental value for the liquid (50.5 mPa.s at 313 K)[42].

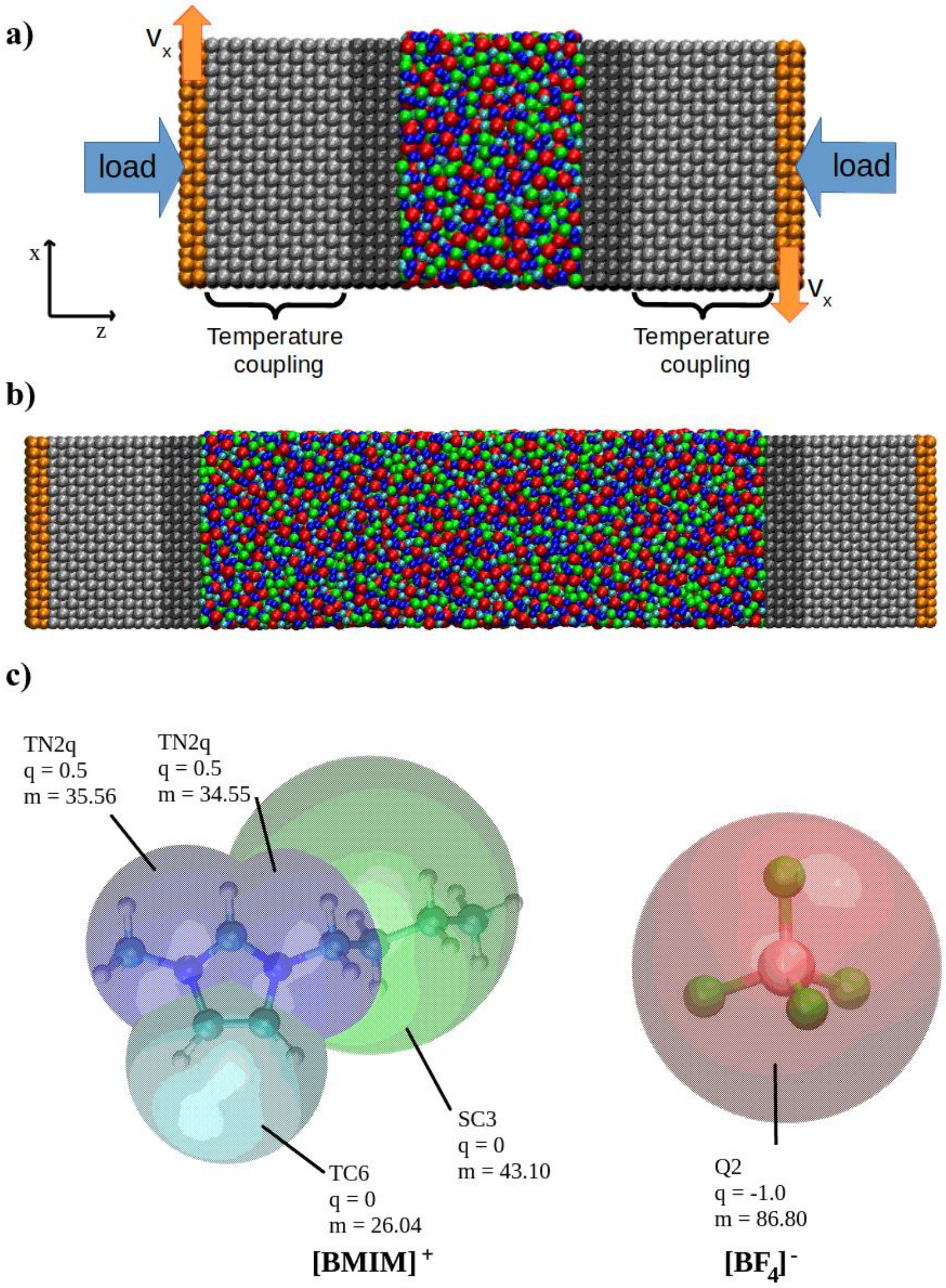

**Figure 1.** a) Graphical representation of the model systems with 1050 ion pairs, with the atoms of the solid surface at which a constant velocity was inforced in x direction showed in orange and the atoms coupled to a thermal bath showed in light gray. b) Similar representation of the system with 4200 ion pairs. c) Structural formula (ball and stick model) superimposed with the corresponding coarse grained model (van der Waals spheres) for the ions indicating the site type in Martini 3.0 (which determines the Lennard-Jones parameters for non-bonded interactions), the charge q in units of elemental charge and the mass m in g/mol.

The solid surfaces were created using the “regular” site size of Martini 3.0 force field with the Lennard-Jones potential (Equation 1) parameter $\sigma$ = 0.47 nm. As a result, the interaction potential minimum between two sites occurs at a distance of 0.528 nm, which was then used as the neighboring sites distance to prepare solid substrates. The solids were generated by replicating a face-centered cubic (FCC) unit cell by 10 times in the cartesian directions $x$ and $y$, and by 9 times in direction $z$, in which the planar interface with the liquid was created with the (100) face exposed (Figure 1). When combined with the IL, this renders a total of 12450 interaction sites in the model with 1050 ion pairs, and 28200 sites in the model with 4200 ion pairs. The boxes were stretched along the $z$ axis to a length of 40 nm in order to create a vacuum on the outside of each solid surface. This spacing ensures that the periodic boundary conditions in the $z$-direction do not introduce any interaction between the solids.

$$U(r) = 4\varepsilon\left[\left(\frac{\sigma}{r}\right)^{12} - \left(\frac{\sigma}{r}\right)^{6}\right] \tag{1}$$

The densities of the liquid and of the solid substrates were relaxed by first running a molecular dynamics simulation with the whole system coupled at 300 K temperature and with pressure coupling only over $x$ and $y$ directions with $p$ = 1 atm. In this step, even while the $z$ dimension of the simulation box is kept constant, both the liquid and the solids can expand in the $z$-direction in opposition to the vacuum produced in the simulation box. The solid substrates were separated into three zones for the production runs, as indicated by the colors in Figures 1.a and 1.b. The two layers closest to the vacuum region made up the first region (orange sites in Figure 1.a). Every atom in this region was forced to move at a constant speed $V_x$ in the $x$ direction while being frozen in the $y$ direction. A constant force was applied in the $z$ direction to provide a constant external load of 1 atm perpendicular to the surface. The central region (light gray) consists of 12 layers and it was coupled to a thermal bath at 315 K using Nosé-Hoover thermostat.[43] Before calculating the temperature, particle velocities were subtracted from the average velocity of each layer caused by the movement imposed on the surfaces, which should only include the disordered thermal motion and not the collective motion caused by the shear. No thermostat was applied in the liquid and in the 4 layers of the solid substrate closer to them, enabling the establishment of a temperature gradient between the liquid and the thermalized region of the solid surface. By using this approach, the shear friction caused by the movement of the surfaces allows the liquid temperature to change.

In order to verify the effect of the chemical nature of the surfaces on the IL, three distinct surfaces were created using the same crystalline structure as mentioned above but varying the interaction parameters with the liquid:

A - the reference surface with similar interaction energy with both the anion and the cation ring;

B - enhanced interaction with anions, which is what one could anticipate from a metallic surface[6,27,37] or an acid ceramic surface with free OH groups capable of donating hydrogen bonds to the anion;

C - enhanced interaction with cation rings that resembles a basic ceramic surface with exposed electronegative atoms that can accept hydrogen bonds from cations.

The interaction parameters between the IL sites and the solid surface sites are given in Table 1. Martini 3.0 force field classifies the interactions between various particle types into several levels, from 21 (weaker interaction) to 0 (stronger interaction).[39] In order to create surface A that provides good adhesion with the IL, it is necessary for the interaction to be relatively strong in order to prevent slip when shear is applied. However, we also want interaction parameters that produce an average interaction energy between the cation and anion rings that is quite similar. This is accomplished by using different levels of interaction between the surface and the cation and the anion sites: interaction level 1 for the anion and interaction level 8 for cation. This yields an average interaction energy of -1234 ± 8 kJ/mol between each surface and the cation rings and -1284 ± 4 kJ/mol between the surface and the anions. If the same interaction level were utilized the adsorption of cations would be strongly preferred since it has more interaction sites. Surface B was created by four times strengthening the anion interaction compared to surface A, whereas surface C involved four times strengthening the cation ring sites interactions (Table 1). For the interaction between surface sites, $\sigma = 0.47$ nm and $\varepsilon = 60$ kJ/mol were used. This is large enough to preclude any dissociation or deformation of the solid surfaces as a result of the interaction with the liquid or of the shear. Simulations of surface A were performed using both the model system with 1050 and the one with 4200 ion pairs, while only the model with 1050 ion pairs was employed for surfaces B and C.

**Table 1.** Lennard-Jones parameters for the interaction between ionic liquid sites (see Figure 1.c) and the surfaces.

| | $\varepsilon$ (kJ/mol) | | | $\sigma$ (nm) |
|---|---|---|---|---|
| Surface | A | B | C | all surfaces |
| TN2q (cation ring) | 2.94 | 2.94 | 11.76 | 0.395 |
| TC6 (cation ring) | 2.94 | 2.94 | 11.76 | 0.395 |
| SC3 (cation tail) | 3.38 | 3.38 | 3.38 | 0.430 |
| Q2 (anion) | 5.49 | 21.96 | 5.49 | 0.470 |

Lennard-Jones and Coulomb interactions were both cut off at 1.1 nm; Coulomb interactions were attenuated using a relative dielectric constant of 5 and long-range effects were taken into consideration using the PPPM (particle-particle - particle-mesh) method, as in our previous work.[16] Periodic boundary conditions were applied in every direction. The shear was created by pushing the solid surfaces to move continuously in opposite directions, as seen in Figure 1. All simulations were performed with LAMMPS software[44] with an integration timestep of 5 fs.

The results presented in section 3.1 employ the larger model system (4200 ion pairs), only the reference surface A, and several values of $V_X$: 4, 5, 8, 10, 15, 20, 30, 40, 45, 50 and 60 m/s. This section concentrates on the effects of the shear rate and viscous heating, and presents two distinct definitions for the viscosity of a thin film. Although surfaces velocities greater than 60 m/s were tested, they resulted in the solid surface slipping, so those simulations will not be discussed here. The sections 3.2 and 3.3 focus on the effect of the chemical nature of the solid surface and will present results for only the smaller system (1050 ion pairs) with surface velocities of 0, 15, 30 and 60 m/s. The simulations with surfaces velocities greater than or equal to 40 m/s were 125 ns long, while simulations at lower velocities were 250 ns long in order to reduce the noise in viscosity calculation at low shear rates. The first 50 ns of each simulation were discarded to ensure the establishment of a steady state. VMD 1.9.3[45] was used to render graphical representations of the structures and to compute radial distribution functions.

# 3. RESULTS AND DISCUSSIONS

## 3.1. Viscosity of a nanoconfined liquid - local and engineering definitions

We calculated the transport properties of the confined liquid film from the molecular dynamics simulations using two different definitions. The first one, which we will refer to as the local definition, involves computing the properties based on variables related to the liquid particles at each slice of the simulation box in the direction $z$ perpendicular to the surfaces. The local shear rate, $\gamma'_l(z)$, is given by the derivative of the average value of the component $x$ of particles velocities, $v_x$, at a box slice at coordinate $z$ in relation to the $z$-axis. The local shear stress, $\tau_l(z)$, is given by the average value of the appropriate component of the pressure tensor at the same slice, and the local viscosity, $\eta_l(z)$, is given by the ratio between the local shear stress and the local shear rate:

$$\gamma'_l(z) = \frac{\partial \langle v_x(z) \rangle}{\partial z} \; ; \; \tau_l(z) = - \langle P_{xz}(z) \rangle \; ; \; \eta_l(z) = \frac{\tau_l(z)}{\gamma'_l(z)}$$

(2)

In our calculations, a width of 0.1 nm was used to define the thickness of the slices perpendicular to the surface, rendering a total of 400 slices across the simulation box, but only slices that have on average at least 25 interaction sites of the liquid inside were included on the analyses, excluding the region outside the liquid film.

The local variables defined in Eq. (2) represent the shear rate, the shear stress and the viscosity that are acting at each distance from the solid surfaces and that control the molecular dynamics of the liquid under shear. In contrast, if a rheometer or other similar device could be built to study a liquid film with only a few nanometers thickness, it would measure (or control) a single value of an engineering shear rate, $\gamma'_e$, given by the relative velocities $V_x$ of the solid surfaces divided by the distance $L$ between them (22 nm for the large model described in this section and 6 nm for the smaller model discussed in sections 3.2 and 3.3), and a single value for an engineering shear stress, $\tau_e$, measured by the ratio between the average force $F_x$ needed to keep the constant movement of the surfaces against the friction exerted by the liquid and the surface area $A_{xy}$ in contact. As a result, a single value of an engineering viscosity, $\eta_e$, is obtained by the ratio of the measured shear stress to shear rate:

$$\gamma'_e = \frac{V_x}{L} \ ; \ \tau_e = \frac{\langle F_x \rangle}{A_{xy}} \ ; \ \eta_e = \frac{\tau_e}{\gamma'_e}$$

(3)

For macroscopic thickness, the local variables (Eq. 2) would converge to the engineering variables (Eq. 3) far from the surface, but for a liquid film of only a few nanometers they can result in very different values. This is shown in Figure 2.a-c for the system with 4200 ion pairs and the reference surface, where local variables are given as a function of $z$ coordinate as the solid curves, while engineering values are given by the horizontal dashed lines of the same color. Note that neither the definitions from Eqs. (2) nor (3) are incorrect or can be regarded as the best definition. Both are suitable for different phenomena. The engineering viscosity determines the force required to move surfaces separated by a thin liquid film and the liquid's resistance to flowing through nanopores or slits, whereas the local viscosity controls the motion of molecules or ions within the liquid.

The local shear rate is lower near to the surfaces than at the interior of the liquid (Fig. 2.a), a result that will be covered in more detail in section 3.3. For small surfaces velocities, the local shear rate achieves a constant value already at 1 or 2 nm away from the surface, corresponding to a linear velocity gradient, but for higher velocities the formation of a plateau is not observed for the film thickness studied, with the local shear rate increasing as one moves away from the surfaces until a maximum is reached in the middle of the liquid film. The local shear rate at the middle point between the surfaces can therefore be significantly higher than the engineering shear rate (dashed horizontal lines). The local shear stress (Fig. 2.b), on the other hand, displays only oscillations close to the surfaces and quickly achieves a constant value equal to the engineering shear stress for any velocity considered here.

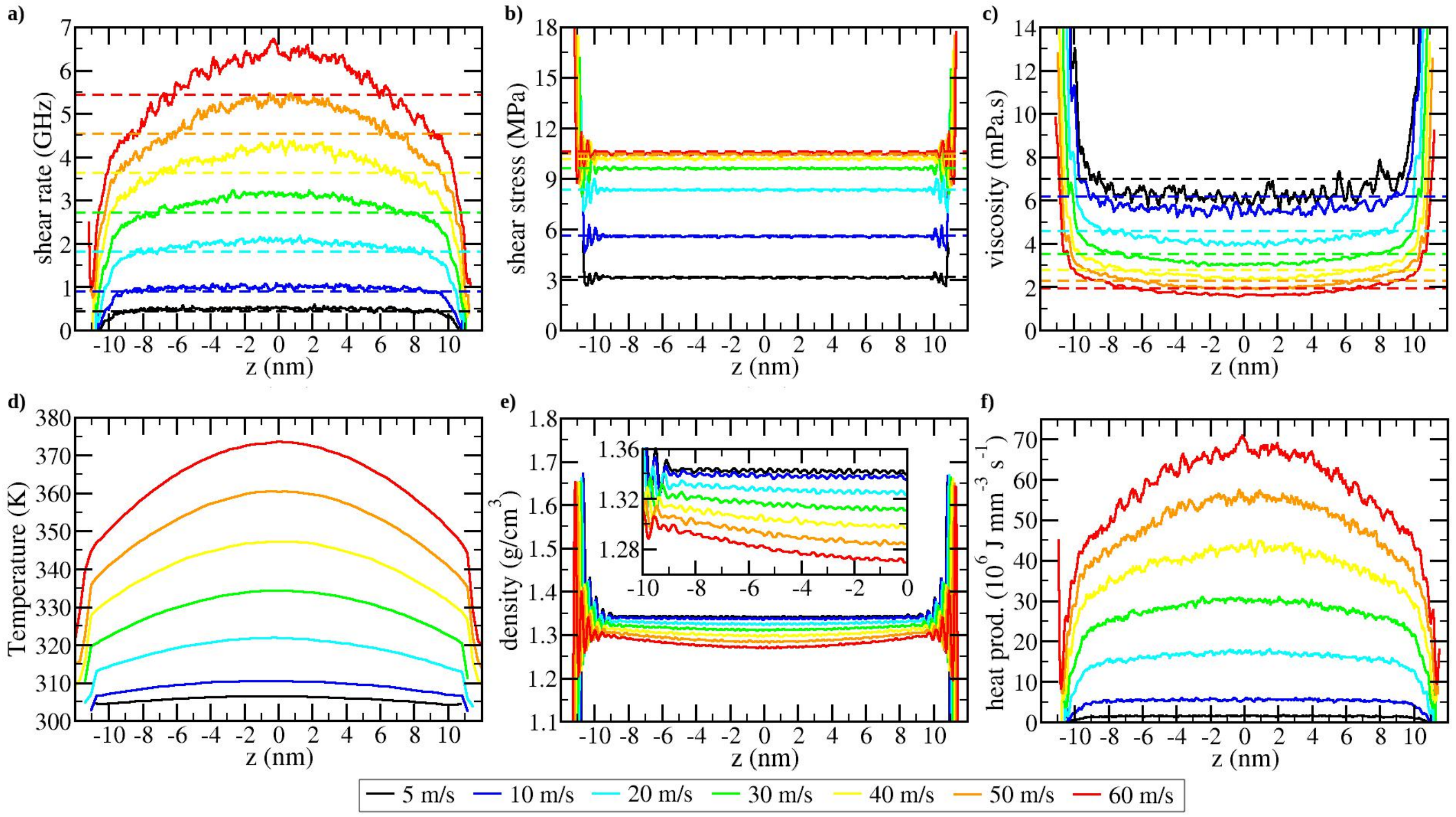

**Figure 2.** Local values of **a)** shear rate, **b)** shear stress, **c)** viscosity, **d)** temperature, **e)** mass density, and **f)** viscous heat production rate inside the ionic liquid along the direction perpendicular to the solid surfaces in the model with surface A and 4200 ion pairs. Different colors indicate various solid surface velocities, as given at the bottom of the figure. The inset in e) displays a zoom for the mass data. In figures a)-c), the horizontal dashed lines indicate the engineering value (Eq. 3), and solid curves indicate the local value (Eq. 2).

As a result, the local viscosity (Fig. 2.c) essentially follows the inverse trend of the local shear rate, with greater values close to the surface and decreasing values toward the bulk. The worse noise-to-signal ratio of local viscosity for small velocities was expected based on earlier non-equilibrium molecular dynamics simulations that produce larger statistical errors for viscosity at relatively small shear rates[15,46] that impose a lower limit to the shear rates that can be properly sampled with a reasonable computational time. As observed for the shear rate, the local viscosity also forms a plateau at the middle of the simulation box for $V_x \leq 10$ m/s, however, the values are still smaller than the zero shear viscosity of 7.97 mPa.s observed in our previous work for the bulk simulations of the same liquid.[16] At the corresponding shear rates the liquid shows a newtonian behavior in bulk simulations,[16] thus the difference here is due to the temperature increase (Figure 2 d) resulting from the viscous heating, an effect that was absent in our bulk simulations performed at a constant temperature of 300 K. The engineering viscosity is halfway between the high value of local viscosity near the surface and the low value in the bulk.

The viscosity (both engineering and local) falls as the shear rate rises. This is in part because of the ionic liquid´s tendency to thin under shear, which was demonstrated in a prior simulation of bulk liquid using both atomistic and coarse-grained models.[15,16,19] The Carreau equation,[47] Eq. (4), can properly describe both the decrease of the engineering viscosity with an increase in the engineering shear rate (red circles with the fit given by the red line in Figure 3) and the variation of the local viscosity as a function of the local shear rate at the middle of simulation box (black squares in Fig. 3). Table 2 lists the best fit parameters. The computed zero shear value of viscosity, $\eta_0$ (dotted lines in Fig. 3), when using the local definition at the center of the liquid film is also smaller than the engineering definition.

$$\eta(\gamma') = \eta_0[1 + (\lambda\gamma')^2]^{(n-1)/2} \quad (4)$$

**Table 2.** Best fit parameters of the Carreau equation, Eq. (4), for the data in Fig. 3 and the correlation coefficient *R* of the fit.

| | $\eta_0$ (mPa.s) | $\lambda$ (ns) | *n* | *R* |
|---|---|---|---|---|
| Local at *z* = 0 | 6.5 | 0.63 | 0.046 | 0.9998 |
| Engineering | 7.3 | 0.71 | 0.058 | 0.9999 |

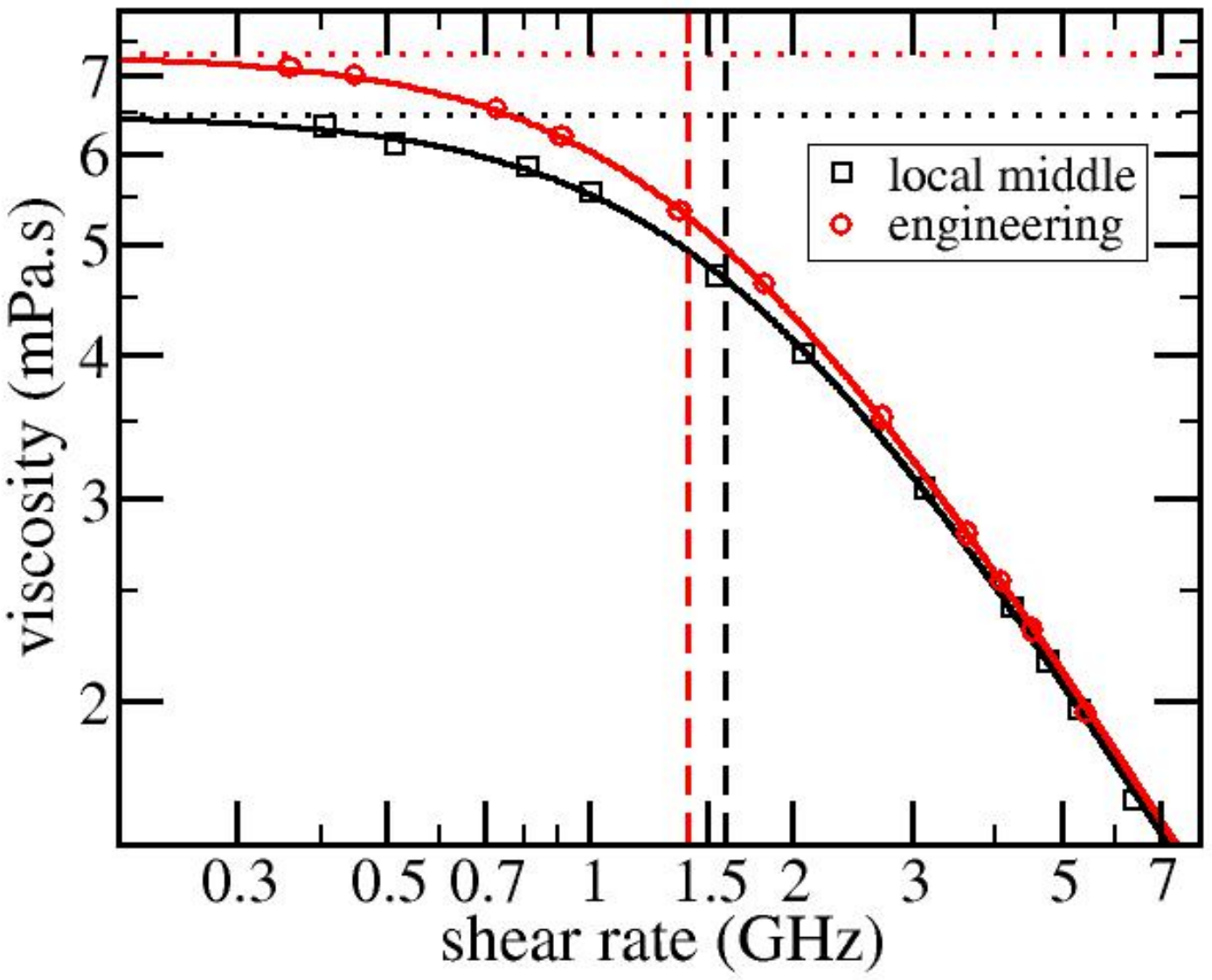

**Figure 3.** Local viscosity calculated at the middle point of the system with 4200 ion pairs as a function of the local shear rate (black), and the engineering viscosity calculated using the average friction force on the surfaces as a function of the engineering shear rate. Solid lines stand for the Carreau equation fit, Eq. (4), horizontal dotted lines indicate the zero shear viscosity, and vertical dashed lines indicate $\lambda^{-1}$.

In addition to the non-newtonian behavior that has previously been observed in simulations at constant temperature,[15,16,19] in this work the temperature of the liquid was allowed to change due to the viscous heating, which results in the increase of liquid temperature with the shear rate (Fig. 2.d), which is also expected to reduce the viscosity of the liquid. Except for very low shear rates, the temperature is not uniformly distributed over the liquid film; instead, it exhibits a maximum value in the middle of the film ($z = 0$). This profile is the result of a balance between the rate of heat production due to the viscous heating and the rate of heat transfer inside the liquid and from the liquid to the solid surfaces, as shown by both experiments,[48] continuous models[49] and atomistic computer simulations.[37] The product of the local shear rate and the local shear stress determines the viscous heat production rate, which is maximal at the center of the liquid film (Fig. 2.f). When the rates of heat transfer and production are equal, a steady state is attained, and even though the temperature is not uniformly distributed throughout the liquid, the temperature at any particular place remains constant.

From the temperature profile and heat production rate per volume $q$, one can compute the thermal conductivity coefficient κ in a system in a steady-state (Equation 5).[50]

$$\frac{d^2T}{dz^2} = -\frac{q}{\kappa}$$

(5)

The numerical calculation of a second-order derivative from the temperature profile introduces larger errors, so instead of the direct calculation, we fit a second-degree polynomial to the temperature profiles from Figure 2.d between $z$=-5 and $z$=5 nm and uses the coefficient of the quadratic term instead. For the q value, due to the noise in the profiles, we also take the average value between $z$=-5 and $z$=5 nm. The results are given in Table 3 and similar results were obtained if we took only the data from $z$=-2 to $z$=2 nm. The thermal conductivity of our model system is lower than the experimental value for [BMIM][BF4] of 0.185 W $m^{-1}$ $s^{-1}$ at 310 K[51] at small shear rates, as expected due to the decreased number of degrees of freedom and weakened interactions in coarse grained models. However, the κ values increase as the shear rate increases until a limit value around 0.27 W $m^{-1}$ $s^{-1}$. This change is not due to temperature increase since the expected effect of temperature is to reduce the thermal conductivity of the IL. This is instead a direct effect of the shear rate as also observed recently in calculations for nanoparticles dispersions subjected to shear.[52]

On one hand, the smaller viscosity of the model results in a smaller viscous heating, which would render a temperature increase due to shear smaller than the real liquid. However, the small number of degrees of freedom in a coarse grained model results also in a smaller heat capacity and, at least in the limit of small shear, in a smaller thermal conductivity, which contributes to a higher temperature increase. Some level of error cancelation is thus expected from these opposite effects. In fact, the magnitude of temperature increase seen in our coarse-grained model is similar to that seen in the atomistic simulations of the same ionic liquid by Ntim and Sulpizi for similar shear rates.[37]

**Table 3.** Calculation of the thermal conductivity coefficient κ for different surfaces velocities $V_x$ for the system with 4200 ion pairs using Eq. 5.

| $V_x$ (m/s) | $d^2T/dz^2$ (K nm$^{-2}$) | $q$ ($10^6$ J mm$^{-3}$ s$^{-1}$) | κ (W m$^{-1}$ s$^{-1}$) |
|---|---|---|---|
| 5 | -0.025 | 1.6 | 0.06 |
| 10 | -0.033 | 5.6 | 0.17 |
| 20 | -0.080 | 17.1 | 0.21 |
| 30 | -0.123 | 29.7 | 0.24 |
| 40 | -0.160 | 42.1 | 0.26 |
| 50 | -0.203 | 54.3 | 0.27 |
| 60 | -0.237 | 65.6 | 0.27 |

Since a constant load is applied on the solid surfaces, the temperature increase allows the liquid to expand against this load, which causes the density to decrease as the surface velocity rises (Fig. 2.e). Since the temperature is not uniform throughout the liquid layer at the steady state, the density also fluctuates, being lower in the center where the temperature is higher. The decrease of the liquid density also contributes to reduce the viscosity as shown in simulations of bulk liquids. The density profile also exhibits peaks close to the surfaces, which is related to the formation of liquid layers, an effect that will be discussed in detail in section 3.2.

The speed at which the viscosity decreases as the shear rate increases is controlled by the parameter n from the Carreau equation, Eq. (4). It is close to 0 for both viscosity definitions (Table 2), which implies that the viscosity is essentially proportional to $\gamma'^{-1}$ for shear rates large enough ($\lambda\gamma' >> 1$) and that the shear stress becomes practically independent of surfaces velocities for values greater than 30 m/s (Fig. 2.b). This is a different outcome from that of our earlier work on the bulk liquid at constant temperature without interaction with surfaces,[16] and which yielded a value of $n$ = 0.47. A smaller value of $n$ indicates a more pronounced reduction in viscosity with the shear rate, demonstrating that, in addition to its natural shear thinning behavior, the increase in temperature and ensuing density reduction also contribute to the reduction in liquid viscosity with the increase in shear rate.

In order to filter the effect of temperature, we performed another set of simulations by holding the surfaces' velocities constant at 30 m/s, but coupling the liquid in the thermostat together with the solid surfaces at three different temperatures: 300, 315 and 330 K, being the later the average temperature achieved in film for this velocity when no thermostat was applied in the liquid (Figure 2 d). The profiles of the local shear rate, local shear stress, local viscosity, temperature, density and viscous heat production across the direction z perpendicular to the surfaces are given in Figure 4 together with the results from the simulation at which no thermostat was applied in the liquid (blue curves). The shear rate is essentially independent of the temperature for a constant surface velocity while the shear stress (both local and engineering) decreases with the temperature, rendering a viscosity reduction. The density and local viscosity curves for the simulation of the liquid coupled at 330 K are very similar to the simulation without thermostat, only displaying a smaller curvature as the temperature profile itself also presents a smaller curvature when the thermostat is used. This also renders the same engineering viscosity for liquid not coupled to a thermostat and the liquid coupled to a thermostat at the average value observed in the uncoupled case. Finally, it is noticeable that the viscosity values for the liquid coupled at 300 K and surface speed of 30 m/s (black curve in Figure 4.c) are only slightly smaller than the liquid with surface speed of 5 m/s without thermostat (black curve in Figure 2.c), showing that the temperature increase is more important for the viscosity reduction than the shear thinning behavior in agreement with the previous observations regarding the smaller value of the parameter $n$ of Carreau equation when the liquid is confined and the temperature is allowed to increase. The same comparison also holds for the liquid density, showing that the most important effect is the temperature increase instead of the shear-thinning behavior in this velocity range.

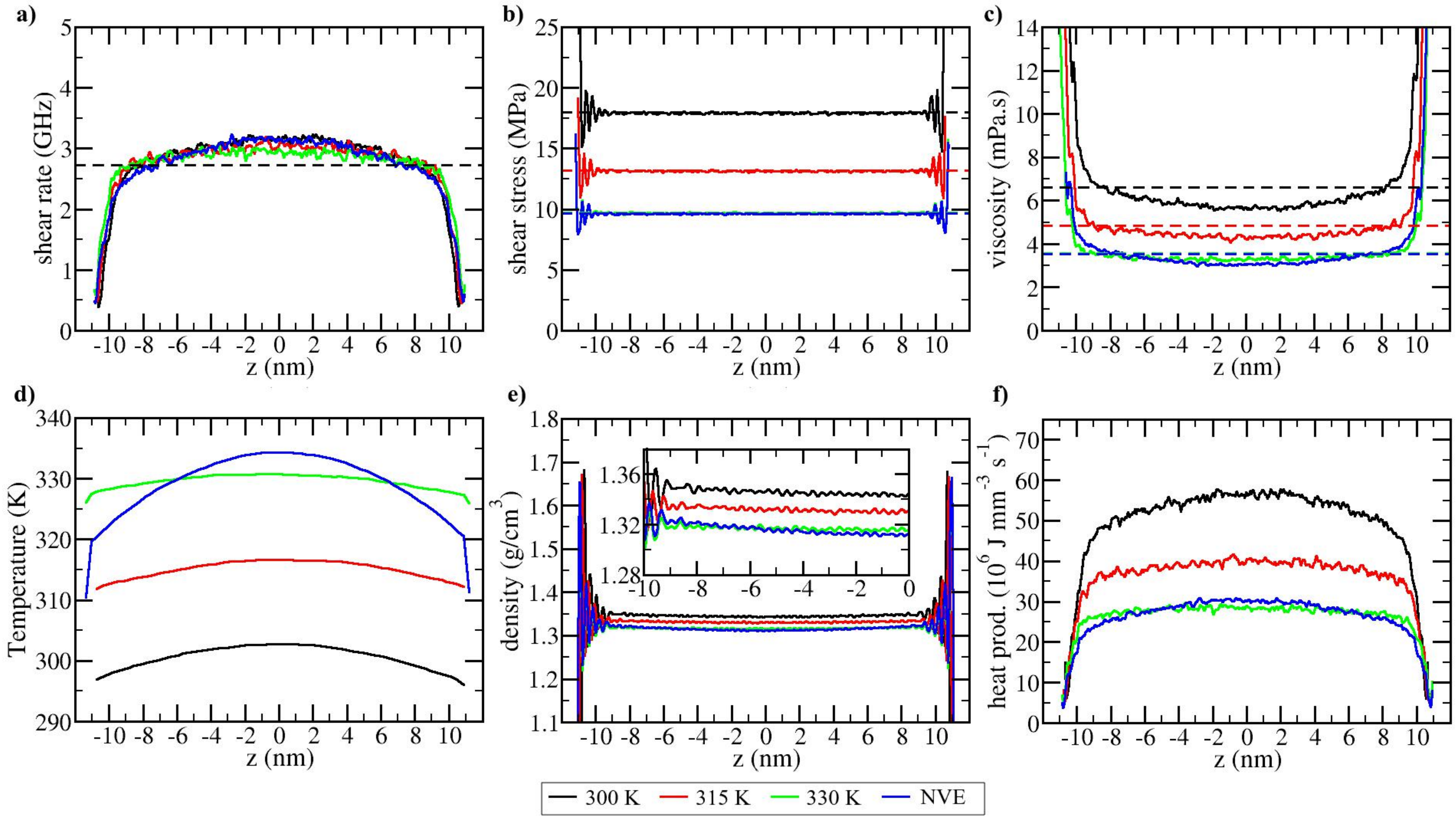

**Figure 4.** Local values of **a)** shear rate, **b)** shear stress, **c)** viscosity, **d)** temperature, **e)** mass density, and **f)** viscous heat production rate inside the ionic liquid along the direction perpendicular to the solid surfaces in the model with surface A and 4200 ion pairs at the surfaces speed of 30 m/s. Blue curves corresponds to the simulations at which no temperature coupling was applied over the liquid, while other colors corresponds to simulations with the whole system coupled at different temperatures showed in the bottom. In figures a)-c), the horizontal dashed lines indicate the engineering value (Eq. 3), and solid curves indicate the local value (Eq. 2).

In the next sections we will concentrate on the effects of the chemical nature of the solid surface, with the discussion in section 3.2 focused on the liquid structure, and the section 3.3 will explore the effects on the liquid viscosity.

### 3.2. Effect of the surface chemical nature on the liquid structure

It is known that the interaction with a flat solid surface causes the introduction of organized layers in a liquid even when a simple hard sphere model without particular interactions is considered. The presence of two species (the cation and the anion) in an IL makes the problem more difficult since distinct interactions with the surface and also specific interactions within the liquid can result in an excess of either cations or anions in a particular layer.[23,28,29,30] The black curves in Figure 5 depict the concentration profile of anion (top row), cation polar ring (middle), and cation apolar tail (bottom) in the absence of shear, while the colored curves depict the same profiles at various surface velocities for the model system with 1050 ion pairs. For the reference surface (A, left column of Figure 5), there is an excess of cation tail sites in contact with the surface (see also the structure of the first layer at Figure 6). This excess is not the product of a particularly strong interaction between the tails and the surface, as the interaction potential with the anion is actually stronger than with the cation tail (Table 1). It is instead a result of the strong interaction between cation rings and anions in the liquid, which tends to leave the aliphatic tails exposed to an interface unless there is a stronger interaction between the surface and the charged species, as happens for surfaces B and C, whose interactions with the anion and cation ring were increased. The concentration of aliphatic tails at the interfaces is significantly reduced as those models of polar solid surfaces are considered, and the layered structure got stronger even at distances as great as 3 nm from the surfaces. The preferred interaction with the anion (surface B) also increases the concentration of cation polar ring at the first layer since cations tend to intercalate between adsorbed anions (structures in Figure 6). For enhanced interaction with the imidazolium ring, the opposite is true (surface C).

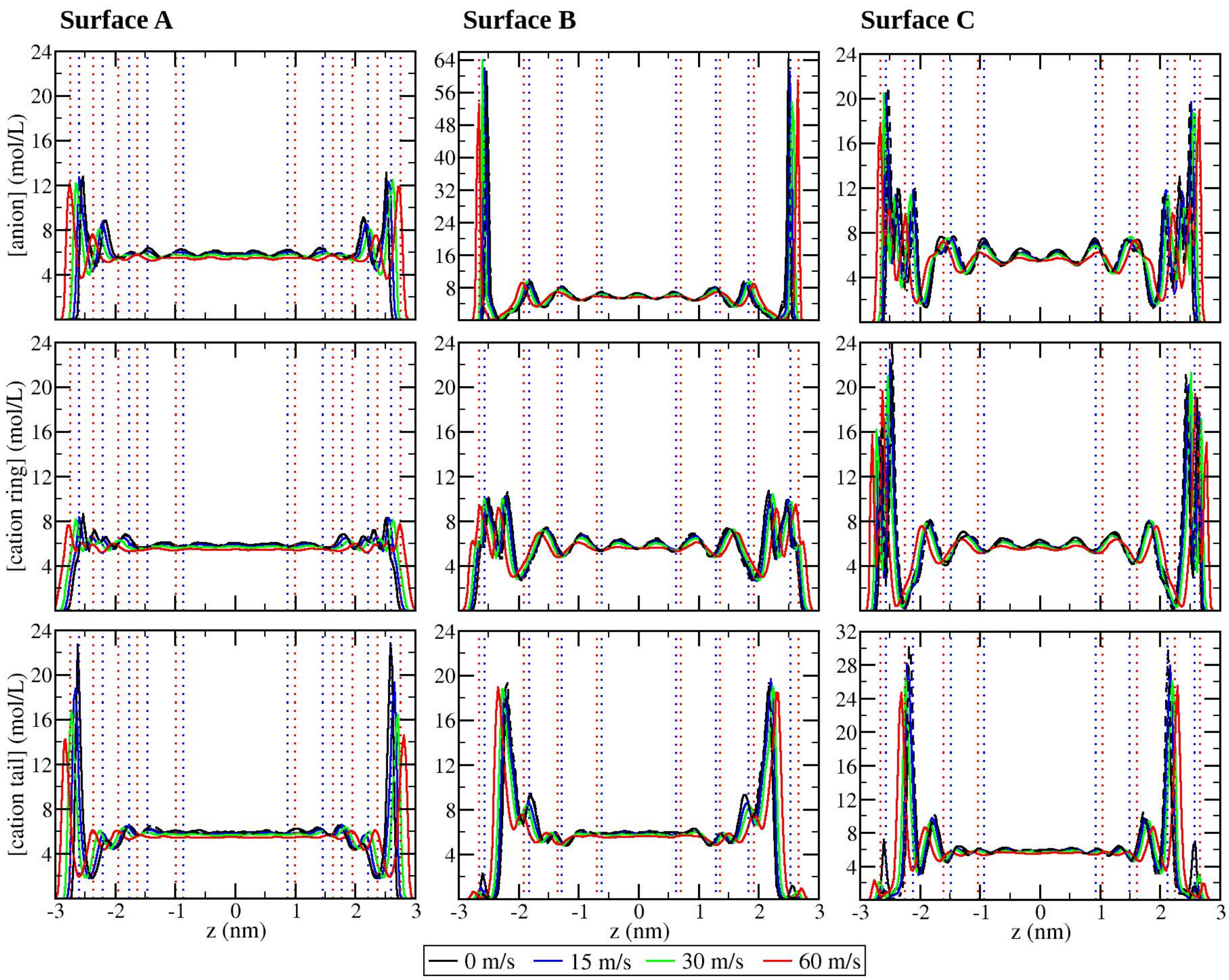

**Figure 5.** Concentration profile of anions, cation rings, and cation tails along the axis perpendicular to the surface. Different columns correspond to different surfaces, and different rows to the different chemical species (from top to bottom: anion, cation ring, and cation tail). Different colors in each panel correspond to the surface speed given at the bottom. Vertical blue and red dashed lines indicate the positions of maximum anion concentration at 15 m/s and 60 m/s, respectively.

Before examining the effect of the shear, it is interesting to investigate how the interaction with the solid surface affects the coordination shells of the IL. Figure 6 shows the radial distribution function, $g(r)$, between the charged sites of the cation and anions at each of the first two layers in contact with the solid surface (dark blue spheres in the structures of Figure 1 and top of Figure 6). Similar data for the anion-anion radial distribution function is shown in Figure 7. The interaction with surface A only slightly intensifies the radial distribution peaks at the liquid layer in contact with the surface, with no appreciable impact on subsequent layers. On the other hand, the layer in direct contact with surfaces B and C develops a solid-like structure when the interaction with the anion or imidazolium ring increases, as seen by the complex $g(r)$ pattern and the graphical representation at the top of Figure 6 and in the insets of Figure 7.

The solid-like character is noticeable for both surfaces B and C, but it is more apparent in the anion-anion radial distribution function at the first layer for surface B, where the anions are arranged in a regular bidimensional square lattice with the distance between anions of 0.7 nm (second anion-anion $g(r)$ peak in Figure 7) and some other anions are intercalated at the center of square lattices (first anion-anion anion-anion $g(r)$ peak; this pattern is more clearly seen in the insets of Figure 7). The anions are packing less regularly on surface C than they are on surface B, presenting small regions with a crystalline-like structure interposed with amorphous liquid-like regions. It is interesting that the crystalline-like regions on surface C display a hexagonal packing of anions instead of the square lattice of surface B. Since the crystalline structure of both substrates is identical, we can conclude that the difference in the interaction strength alone can introduce distinct intermolecular structures within the liquid. Anions pack more tightly because they have larger symmetry than cations, which makes regular packing over the surface easier. This should be valid for symmetric anions like tetrafluoroborate, $[BF_4]^-$, and hexafluorophosphate, $[PF_6]^-$, but not for low symmetry anions like bis-trifluoromethylsulfonylimide, $[(CF_3SO_2)_2N]^-$. Indeed, atomistic simulations revealed a stronger organization for the IL with $[BF_4]^-$ than for a similar liquid with $[(CF_3SO_2)_2N]^-$ when confined between silica surfaces.[29] The second and further layers from surfaces B and C exhibit $g(r)$ profiles typical of the liquid state and qualitatively similar to those seen for surface A, with only small differences in their amplitudes.

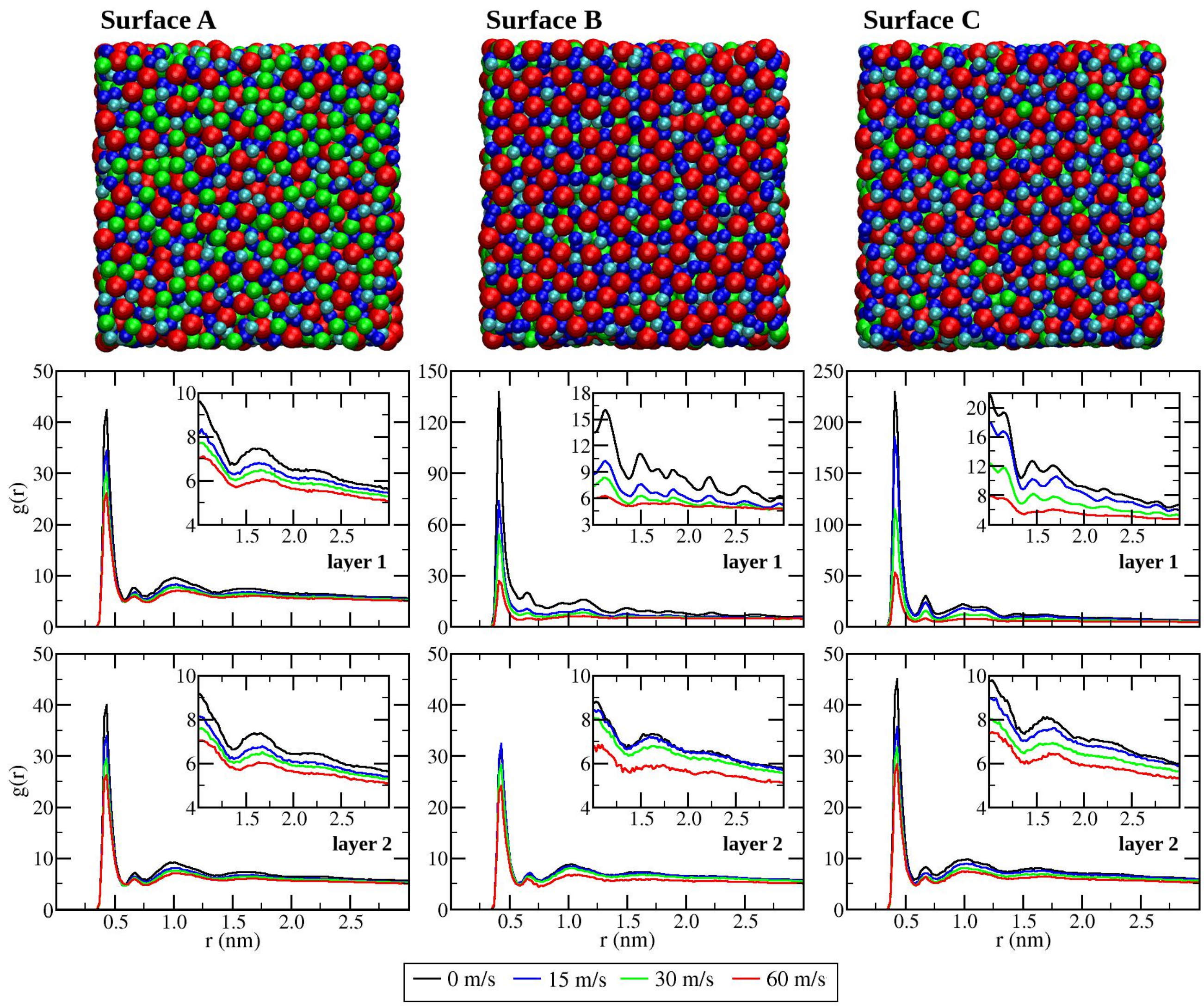

**Figure 6.** Top: Graphical representation of the liquid in contact with each surface, with anions denoted by red, the charged and uncharged sites of the cation ring in blue and cyan, respectively, and the tail sites of the cation in green. Bottom: Radial distribution function between the cation charged sites and anions at the two first layers parallel the solid surfaces. Different columns correspond to different surfaces, while different rows correspond to different layers, being layer 1 (top row) in contact with the surface. Different colors denote different surfaces speed. Each panel's insets zoom out to show the long-range structure.

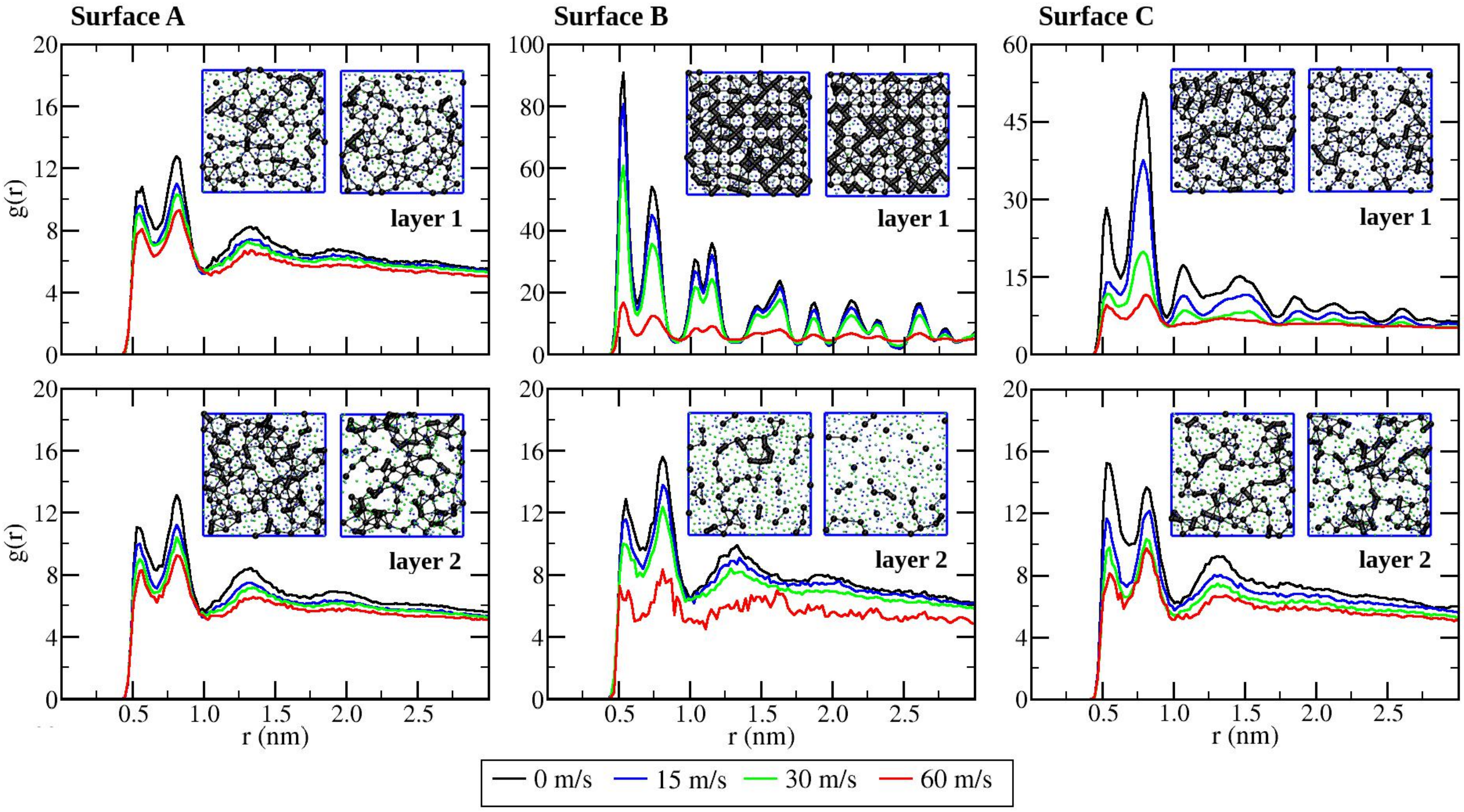

**Figure 7.** Radial distribution function between anions at the first two layers parallel to the solid surfaces. Different columns correspond to different surfaces, while different rows correspond to different layers, being layer 1 (top row) in contact with the surface. Different colors denote different surfaces speed. Insets at each panel depict instantaneous representations of the anions of the respective layer as black spheres in absence of shear (left structures) and at the surfaces speed of 60 m/s (right structures). Anions in immediate contact were connected by thicker lines (first $g(r)$ maximum), whilst second neighbors were connected by thinner lines (second $g(r)$ maximum). Smaller spheres represente the cation sites at the same layer.

The shear does not fully destroy the coordination shells of the liquid (colored curves in Figures 6 and 7) or the liquid layers generated by the solid surface (colored curves in Figure 5), but it does make them more diffuse. This phenomenon is comparable to the partial breakdown of coordination shells seen in non-equilibrium simulations of bulk ionic liquids, which was linked to a decrease in liquid viscosity with an increase in shear rate.[15,16,19] Here, only the solid surfaces were coupled to a thermostat, so that the temperature of the liquid increases with shear rate as a result of viscous heating, reaching 397 K, 368 K, and 372 K at the middle of the liquid film for surfaces A, B, and C, respectively, at maximum shear rate. Additionally, when the temperature rises, the liquid can expand in opposition to the solids' continuous load. As a result, the liquid's density falls while its shear rate rises. Hence, the reduction in the liquid structure that is seen on this work as the surface velocity increases is caused by a combination of the shear, temperature increase, and density decrease. The liquid layer in contact with surface C experiences partial destruction of the crystalline domains at 60 m/s (red curves), but not the liquid in contact with surface B, where the intensity of the $g(r)$ peaks decreases but the profile remains the same. The reduction of the structure in contact with surface B is thus similar to what would be expected when heating a crystal without a phase transition occurring in the temperature range. This distinction also results from the ease with which high-symmetric anions can be packed over surfaces as opposed to low-symmetric cations.

### 3.3. Effect of the surface chemical nature on transport properties

Considering the reference surface A, both the local shear rate and shear stress exhibit local maxima that coincide with the maximum of anion concentration at the first layer and both pass through a minimum between the first and the second liquid layers (Figure 8, where the positions of the maxima of anion densities from Figure 5 are shown by vertical dotted lines for 15 and 60 m/s). There is a higher local viscosity at the liquid layer in contact with the surface because the effect on the stress is substantially stronger than the effect on the rate at the first layer close to surface A. The more organized coordination shells of the IL in contact with the solid surface provide justification for the higher local viscosity as suggested by previous works regarding atomistic simulations of bulk liquids.[19] As one moves away from the surfaces, the shear stress exhibits a few further oscillations that coincide with the maximum and minimum of density profiles and tends swiftly to a constant value, as observed also in the larger model covered in section 3.1. On the other hand, the local shear rate continues to rise as moving toward the middle of the liquid film, indicating a reduction in the local viscosity as one moves away from the surfaces. If a linear velocity profile were established inside the liquid, the local shear rate should be constant and equal to the engineering shear rate for surface A. However, the interaction with the surfaces, and the resulting layered structure, implies that the local shear rate close to the surfaces is smaller than the engineering value (dashed horizontal lines), but becomes larger a few layers away from the surfaces.

The local viscosity exhibits the opposite pattern, with a minimum at the intermediate point between the surfaces where the temperature exhibits the largest value (see section 3.1). The difference between the local and the engineering values would disappear far from the surfaces if the film thickness is macroscopic, as can be seen by the variations between engineering and local viscosity at the center of the liquid, which are larger for the small systems presented here than for the systems with four times thicker films of section 3.1.

As expected, an increase in surface velocity (shown in different colors in Figure 6) causes an increase in the local shear rate and the engineering shear rate, and a consequent decrease in the local and engineering viscosities. These changes are caused by the liquid's shear thinning behavior, which has been observed in previous work,[15,16,19] as well as an increase in temperature brought on by viscous heating, which was covered in section 3.1. However, because the distance between the surfaces is nearly four times closer in this smaller model system, the shear rate is around four times higher for the same surface velocity $V_x$. This results in stronger viscous heating and higher temperatures as the separation between the solid surfaces diminishes, with the temperature at the center of the simulation box reaching 397 K for the smaller model as opposed to 374 K in the bigger model for $V_x$ = 60 m/s. In summary, the reduction of the film thickness at a constant load and surface viscosities results in a higher shear rate, which produces a higher viscous heating and a higher temperature increase inside the liquid. The higher temperature decreases the viscosity of the liquid under shear (both local and engineering) as can be observed by comparing the viscosity data for the larger system (Figure 2 c) with the one for the smaller system with surface A (Figure 8). Comparing the results for L=6 nm from this section with the ones for L=22 from section 3.1 show that the local viscosity at the middle of the simulation box is closer to bulk values in the later, especially for the smaller shear rates at which the liquid viscosity displays newtonian behavior and the temperature increase due to viscous heating gets smaller.

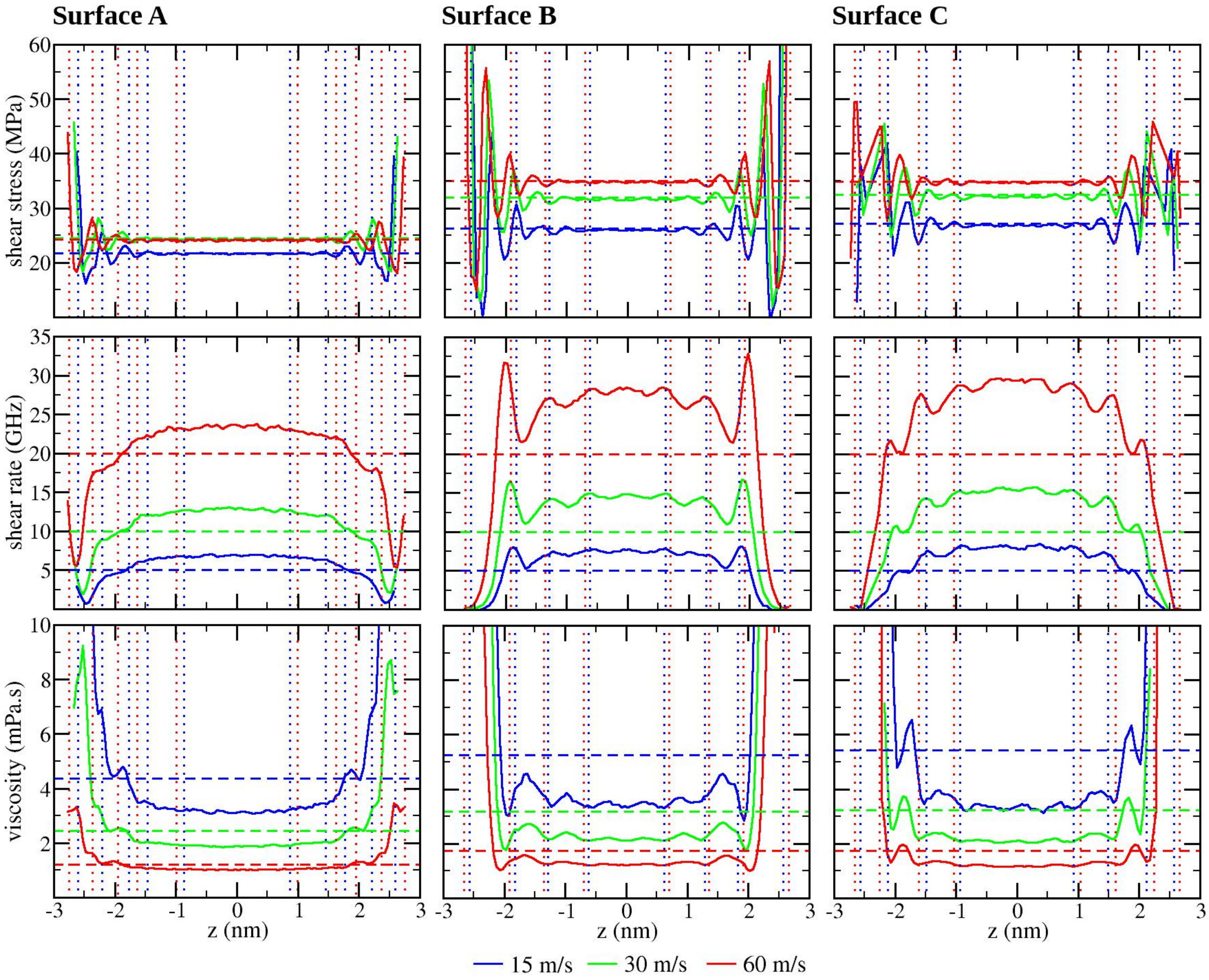

**Figure 8.** Local shear stress (top), shear rate (middle), and viscosity (bottom) in the direction perpendicular to the solid surfaces. Different columns correspond to different surfaces; different colors correspond to different surfaces speed. Vertical blue and red dashed lines indicate positions of maximum anion concentration at 15 m/s and 60 m/s, respectively. Horizontal dashed lines indicate the engineering values of the respective properties for each surface and speed using the same colors as the local value curves.

The layered structure gets stronger for the surfaces with enhanced interactions (B and C), as previously described, and has bigger influence on the local shear stress and the local shear rate than observed for the surface A. Particularly, the local viscosity at the first layer for these surfaces is ill-defined because the first layer of the liquid travels along with the solid surface, making the local shear rate equal to zero. As was previously observed, while calculating the distance $L$ between surfaces at the engineering shear rate, Eq. (3), the layer in contact with surfaces B and C assumes a solid-like structure and should be regarded as part of the solid surfaces. However, assuming no prior knowledge of the solid-like structure of the liquid at the first layer and not deducting its thickness from the distance $L$, the engineering shear rate is significantly underestimated and the local shear rate far from the surfaces is significantly higher than the expected by the engineering value, more so than for the reference structure where the first layer is sheared. As a result, if the liquid interacts strongly with the substrate, the engineering viscosity can be significantly larger than the local viscosity of the liquid under nanoconfinement. This effect will become negligible as the distance $L$ between the solid surfaces is substantially greater than the thickness of the liquid layers that adquires a solid like structure due to the interaction with the surfaces.

For surface B, the local shear rate exhibits local maxima at each peak of anion concentration (indicated by the vertical dotted lines in Figure 8), with the exception of the first layer, where there is no shear. This produces minimum local values for the viscosity at each anion layer intercalated with maximum values between layers, resulting in a faster movement inside the layers than between the layers. Except for the second layer, surface C exhibits the same effect. This difference is likely due to the weaker packing of ions near the surface, which exhibits higher interaction with cations. While the engineering viscosity is higher for the surfaces with stronger interactions, it is noteworthy that the local viscosity at the center of the liquid film is virtually the same for all surfaces when comparing the viscosity values for the three surfaces. This implies that the strength of the interaction with the surfaces affects resistance to surface movement, to liquid flow, as well as the motion of the molecules close to the surfaces, but did not affect the molecular motion a few layers away from the surfaces. This was also expected from the previous observation that the surface primarily affects the liquid structure at the first layer with only minor effects over further layers discussed in section 3.2.

## 4. CONCLUSIONS

In this paper, two distinct definitions for the viscosity of a nanoconfined liquid film were proposed and calculated: A local definition based on the local pressure tensor and velocity gradient inside the liquid, and an engineering definition based on the friction force measured on the solid surfaces and the surfaces velocities and separation. The local viscosity controls the molecular dynamics of the liquid and is higher than the engineering viscosity close to the interfaces.

A solid-like packing of the ions in the first layer over the surface can result from increasing the intermolecular interactions of the surface with either the anion or the cation; the tendency to form bidimensional crystalline domains is stronger when the interaction with the ion with higher symmetry is increased. The layers with crystalline domains are not sheared; rather, they move together with the solid surface. As a result, the engineering shear rate is underestimated and the engineering viscosity is correspondingly larger than the local viscosity. The complex patterns with maximums and minimums produced by the local viscosity profiles for surfaces with strong interactions with the ionic liquid are associated with the layered structure in the liquid brought about by the presence of the solid surface. As the shear rate increases, both the local and the engineering viscosities decreases because of the non-newtonian behavior of the ionic liquid and, more important, the rise in temperature brought on by viscous heating, which is a behavior that can be represented by the Carreau equation.

The effects of confinement, shear, and interaction between liquid films and solid surfaces discussed here should be extended to other liquid systems besides ionic liquids. Even though the local structure induced by the interaction with the surface does not produce a perceptible effect for macroscopic amounts of liquid as in typical rheology experiments, regarding only the fact that there is no slip between the surface and the liquid, these effects are significant in the context of lubricant thin films over surfaces subject to shear as well as the flow of liquid over nanostructured materials like zeolites and nanoporous filters. The different values of local viscosity at various regions within the nanoconfined liquid as well as their dependence on the chemical nature of the solid substrate can also be important to understand the kinetics of chemical reactions catalyzed by porous materials and the surface reactions involved in the formation of tribological films.

## ACKNOWLEDGMENTS

We are indebted to FAPESP (grants 2017/12063-8, 2019/04785-9, and 2016/21070-5). MCCR is also indebted to CNPq (grant 303045/2021-3). We also thank the “Laboratório Nacional de

Computação Científica (LNCC/MCTI, Brazil)" for the use of the supercomputer SDumont (https://sdumont.lncc.br).

**Data availability statement**

The input files needed to reproduce the simulations performed as well as any other inquire can be directly requested from the corresponding author.

**Author Contribution**

KB: General ideas, literature revision, construction of the model system, execution and analyses of the computer simulations, text writing and revision. MCCR: General ideas, literature revision, text revision, project supervisor, and responsible for infrastructure.

**Conflict of Interest**

The authors declare that the research was conducted in the absence of any commercial or financial relationships that could be construed as a potential conflict of interest.